\documentclass[conference]{IEEEtran}

\usepackage{ifpdf}
\usepackage{cite}

\ifCLASSINFOpdf
  \usepackage[pdftex]{graphicx}
  \usepackage{dblfloatfix}
  \graphicspath{{figures/}}
  \DeclareGraphicsExtensions{.pdf,.png,.jpg}  
\else

\fi

\usepackage{booktabs}  
\usepackage{amsmath}
\usepackage{amssymb}
\usepackage{color}
\usepackage{algorithm}
\usepackage{algpseudocode}
\usepackage{array}

\ifCLASSOPTIONcompsoc
 \usepackage[caption=false,font=normalsize,labelfont=sf,textfont=sf]{subfig}
\else
 \usepackage[caption=false,font=footnotesize]{subfig}
\fi

\usepackage{url}

\begin{document}

\title{An Adaptive Gradient Clipping and Noise Injection Mechanism for Differentially Private Federated Learning}

\author{\IEEEauthorblockN{Wenjing Wei}
\IEEEauthorblockA{Université Paris Cité\\
wenjing.wei2026@outlook.com}
\and
\IEEEauthorblockN{Alla Jammine}
\IEEEauthorblockA{Université Paris Cité\\
alla.jammine@u-paris.fr}
\and
\IEEEauthorblockN{Farid Nait-Abdesselam}
\IEEEauthorblockA{Université Paris Cité\\
farid.nait-abdesselam@u-paris.fr}}

\maketitle

\begin{abstract}
Differentially private federated learning must balance privacy protection against model accuracy and training efficiency. Static gradient clipping applies a fixed threshold throughout training and across model layers, which can cause excessive clipping when the threshold is too small or unnecessarily large noise when it is too large. This paper presents DDP-SA-adaptive, an adaptive gradient clipping and noise adding mechanism for differentially private federated learning with secure aggregation. At each communication round, every client determines a separate clipping threshold for each model layer from the median of its per-sample gradient norms. The resulting layer-wise thresholds adapt to the evolving gradient distributions and calibrate the Laplace noise added before the updates are encoded and secret-shared among intermediate aggregation servers. We evaluate the proposed mechanism on a federated regression task in terms of efficiency, accuracy, privacy, convergence, clipping norm, and noise magnitude. Compared with the static DDP-SA baseline, DDP-SA-adaptive reduces the number of communication rounds by 6.81\%, total training time by 19.21\%, and average per-round training time by 13.33\%, leading to improved training efficiency. It also reduces test loss by 98.74\% and increases test $\mathrm{R}^{2}$ by 3.41\%, leading to improved model accuracy. To attain $\mathrm{R}^{2}=0.99$, the adaptive mechanism operates with a privacy budget of approximately $\epsilon=0.1$, compared with $\epsilon=0.4$ for static DDP-SA, thus providing stronger privacy protection and achieving stronger privacy guarantees. These results demonstrate that round-wise, layer-wise adaptation can improve the privacy-accuracy-efficiency trade-off of differentially private federated learning.
\end{abstract}


\IEEEpeerreviewmaketitle 

\section{Introduction}
Machine learning has become a core technology for intelligent applications in computer vision, speech recognition, and natural language processing. In conventional centralized training, however, raw user data must be uploaded to a remote server, which raises serious privacy concerns when the data contain sensitive personal information~\cite{yang2019federated}. Federated learning (FL) mitigates this risk by keeping private data on local devices and exchanging only model updates, such as parameters or gradients, so that multiple clients can collaboratively train a shared global model without centralizing their datasets~\cite{mcmahan2017communication}. Despite this design, recent studies have shown that shared updates can still leak private information through inference attacks, including membership inference, attribute inference, and gradient inversion~\cite{nasr2019comprehensive,kairouz2021advances,zhu2019deep}.

Differential privacy (DP) and secure multi-party computation (MPC) are two principal defenses against such attacks. DP protects individual records by injecting calibrated noise into local updates before they leave the client~\cite{dwork2014algorithmic}, but larger noise typically degrades model accuracy. MPC-based secure aggregation allows a server to obtain only the sum of client updates without observing any individual contribution~\cite{bonawitz2017practical}, yet cryptographic protocols often incur substantial communication and computation overhead. Hybrid frameworks that combine local DP with secure aggregation, such as distributed differential privacy through secure aggregation (DDP-SA)~\cite{wei2026ddpsa}, reduce the noise each client must add while still preventing the parameter server from inspecting individual updates. In such pipelines, gradient clipping is indispensable: it bounds the sensitivity of each update so that Laplace or Gaussian noise can be calibrated to a target privacy budget.

Existing DP-FL systems, including static DDP-SA, typically employ a fixed clipping threshold for the entire model and for all communication rounds. A threshold that is too small clips informative gradients and slows convergence, whereas a threshold that is too large inflates the sensitivity and forces the client to inject excessive noise. Moreover, different layers of a neural network generally exhibit different gradient norm scales, so a single global threshold cannot match the local gradient distribution of every layer. Adaptive clipping methods have been studied in centralized and federated DP training~\cite{andrew2021differentially,mcmahan2023privacy,fu2022adap}, yet they are not directly tailored to a distributed DP setting in which noisy updates are subsequently encoded and secret-shared among intermediate aggregation servers.

To address this gap, we propose DDP-SA-adaptive, an adaptive gradient clipping and noise injection mechanism for differentially private federated learning with secure aggregation. At each communication round, every client estimates a separate clipping threshold for each model layer from the median of its per-sample gradient norms, clips the corresponding layer gradients, and injects Laplace noise whose scale is calibrated to the adapted threshold and the allocated privacy budget. The noisy updates are then fixed-precision encoded and secret-shared for secure aggregation, preserving the end-to-end privacy pipeline of DDP-SA while adapting the clipping-noise trade-off to the evolving training dynamics.

The main contributions of this paper are summarized as follows:
\begin{enumerate}
    \item We design a round-wise and layer-wise adaptive clipping mechanism for DP-FL with secure aggregation, in which each client sets the clipping threshold of each layer to the median of its current per-sample gradient norms.
    \item We couple the adaptive thresholds with Laplace noise injection and per-round, per-layer privacy budget allocation, so that the noise scale tracks the local sensitivity without changing the secure aggregation workflow.
    \item We empirically evaluate DDP-SA-adaptive against static DDP-SA on a federated regression task in terms of efficiency, accuracy, privacy, convergence, clipping norm, and
    noise magnitude. The results show fewer communication rounds, shorter training time, lower test loss, higher test $\mathrm{R}^{2}$, and a smaller privacy budget for the same accuracy target.
\end{enumerate}

The remainder of this paper is organized as follows. Section~II reviews related work on privacy-preserving federated learning, differential privacy, and adaptive clipping. Section~III presents the system model and the proposed DDP-SA-adaptive mechanism. Section~IV reports the experimental results. Section~V concludes the paper.

\section{Related Work}
Federated learning enables collaborative model training without centralizing raw data and has been widely adopted in privacy-sensitive domains such as healthcare, finance, and the Internet of Things~\cite{mcmahan2017communication,yang2019federated}. Nevertheless, exchanging model parameters or gradients remains vulnerable to privacy inference attacks. Prior work has shown that shared updates can leak sensitive attributes~\cite{melis2019exploiting}, enable adversarial reconstruction via generative models~\cite{hitaj2017deep}, or recover training samples from gradients~\cite{zhu2019deep,geiping2020inverting}. Membership inference and related threats further demonstrate that FL alone does not provide adequate privacy guarantees~\cite{nasr2019comprehensive,kairouz2021advances}. These attacks motivate stronger defenses based on differential privacy, cryptography, or their combination.

Differential privacy has become a standard tool for protecting FL updates~\cite{dwork2014algorithmic,wei2020federated,fu2024differentially}. Client-level and local DP mechanisms inject calibrated noise into gradients or parameters before aggregation, thereby limiting the influence of any individual record. Because noise magnitude is determined by sensitivity, gradient clipping is a necessary preprocessing step. Most early DP-FL systems use a fixed clipping threshold throughout training. Such a static threshold is difficult to tune: an overly small value discards useful signal, whereas an overly large value inflates sensitivity and forces excessive noise. Recent studies therefore explore adaptive noise or clipping schedules to improve the privacy-utility trade-off. Fu et al.~\cite{fu2022adap} and Liu et al.~\cite{liu2022adaptive} adapt the noise scale during federated training, while Andrew et al.~\cite{andrew2021differentially} and McMahan et al.~\cite{mcmahan2023privacy} adapt clipping thresholds online. Adaptive optimization and noise-aware training have also been investigated for non-IID and heterogeneous federated settings~\cite{chen2024differentially,malekmohammadi2024noise}. These approaches improve utility over naive static DP-FL, but they are typically designed for central or local DP pipelines in which the server observes individual (noisy) updates, rather than for a distributed DP setting with subsequent secure aggregation.

Cryptographic techniques provide a complementary line of defense. Secure aggregation based on MPC allows the server to learn only the sum of client updates~\cite{bonawitz2017practical}, while homomorphic encryption supports encrypted aggregation or training~\cite{aono2017privacy}. Hybrid schemes combine DP with secure computation to obtain both statistical privacy and cryptographic confidentiality~\cite{kairouz2021distributed}. In particular, DDP-SA integrates client-side Laplace perturbation with full-threshold additive secret sharing across intermediate servers, so that the parameter server reconstructs only an aggregated noisy update~\cite{wei2026ddpsa}. This design strengthens end-to-end privacy relative to LDP or MPC alone. However, the original DDP-SA pipeline still relies on a single static clipping threshold applied to the entire gradient vector in every round. As a result, it cannot track the contraction of gradient norms during training, nor the heterogeneous gradient scales across neural network layers.

Overall, existing work either adapts clipping and noise outside a secure aggregation pipeline, or combines DP with secure aggregation under static sensitivity control. DDP-SA-adaptive addresses this gap by embedding round-wise, layer-wise median-based clipping and correspondingly calibrated Laplace noise into the DDP-SA workflow. The adaptive thresholds remain local to each client, while the noisy updates continue to be encoded and secret-shared for secure aggregation, thereby improving the privacy-accuracy-efficiency trade-off without altering the aggregation architecture.

\section{Method of Research}
This section presents the system model and threat model, the proposed DDP-SA-adaptive mechanism, and the associated privacy analysis. The overall workflow follows the three-layer DDP-SA architecture of clients, intermediate servers, and a parameter server~\cite{wei2026ddpsa}. The key difference is that each client replaces the static clipping threshold with round-wise, layer-wise median-based clipping and correspondingly calibrated Laplace noise.

\subsection{System Model and Threat Model}
We consider $n$ clients $C=\{C_1,\dots,C_n\}$, $m$ intermediate servers $S=\{S_1,\dots,S_m\}$, and one parameter server (PS). Client $C_i$ holds a private local dataset $D_i$ of size $N_i$ and never uploads raw samples. The global model is parameterized by $\theta=\{\theta^{(1)},\dots,\theta^{(L)}\}$, where $\theta^{(l)}$ denotes the parameters of layer $l\in\{1,\dots,L\}$. Let $N=\sum_{i=1}^n N_i$ denote the total number of training samples.

Unlike conventional two-party FL with only clients and a PS, the intermediate server layer enables secure aggregation via full-threshold additive secret sharing (ASS). At each communication round, every client first privatizes its local update with calibrated Laplace noise, encodes the noisy update with fixed precision encoding, and decomposes the encoded vector into $m$ additive shares. Each intermediate server receives one share per client, sums the shares it holds, and forwards only the partial aggregate to the PS. The PS reconstructs the global aggregate after collecting all $m$ partial sums, updates the model, and broadcasts the new parameters to the clients.

We adopt a semi-honest (honest-but-curious) adversary model: clients, intermediate servers, and the PS follow the protocol correctly, but may attempt to infer private information from observed messages. We assume authenticated channels and that an adversary may corrupt at most $f<m$ intermediate servers. Under full-threshold ASS, any strict subset of shares is information-theoretically independent of the secret. Confidentiality degenerates only if an adversary obtains all $m$ shares, which is an inherent limitation of full-threshold secure aggregation~\cite{bonawitz2017practical}. The security goal is that no individual client update is revealed in transit or at the PS; the PS observes only the reconstructed aggregate of already privatized updates.

\subsection{Problem Setup}
At round $t$, client $C_i$ receives the current model $\theta_t$ and computes per-sample gradients
\begin{equation}
g_t^{(l)}(x)=\nabla_{\theta_t^{(l)}}\mathcal{L}(\theta_t,x)
\end{equation}
for each layer $l$ and each local sample $x\in D_i$. In static DDP-SA, a single fixed clipping threshold $\Delta$ is applied to the entire gradient vector in every round. If $\Delta$ is too small, informative gradients are over clipped; if $\Delta$ is too large, sensitivity is overestimated and the Laplace noise becomes unnecessarily large. Moreover, different layers typically exhibit different gradient norm scales, so one global threshold cannot match every layer. DDP-SA-adaptive therefore determines a separate clipping threshold for each client, round, and layer from local gradient statistics.

\subsection{Adaptive Layer-wise Clipping}
For client $C_i$, round $t$, and layer $l$, let
\begin{equation}
\mathcal{G}_{t,i}^{(l)}=\bigl\{\|g_t^{(l)}(x)\|_1:x\in D_i\bigr\}
\end{equation}
denote the multiset of per-sample $\ell_1$ gradient norms. The adaptive clipping threshold is defined as
\begin{equation}
\Delta_{t,i}^{(l)}=\operatorname{median}\bigl(\mathcal{G}_{t,i}^{(l)}\bigr).
\end{equation}
Each per-sample gradient is then clipped by
\begin{equation}
\overline{g}_t^{(l)}(x)
=
\frac{g_t^{(l)}(x)}{\max\!\left(1,\dfrac{\|g_t^{(l)}(x)\|_1}{\Delta_{t,i}^{(l)}}\right)},
\end{equation}
which guarantees $\|\overline{g}_t^{(l)}(x)\|_1\le\Delta_{t,i}^{(l)}$. The client forms the layer-wise summed gradient
\begin{equation}
G_{t,i}^{(l)}=\sum_{x\in D_i}\overline{g}_t^{(l)}(x).
\end{equation}
Applying this procedure independently to every layer yields thresholds $\Delta_{t,i}^{(1)},\dots,\Delta_{t,i}^{(L)}$ that typically decrease as training progresses and gradient magnitudes contract.

\subsection{Adaptive Noise Injection and Secure Aggregation}
Let $\epsilon_{\mathrm{total}}$ denote the total privacy budget. We first allocate a per-round budget $\epsilon_t=\epsilon_{\mathrm{total}}/T$ and then split it uniformly across layers,
\begin{equation}
\epsilon_t^{(l)}=\frac{\epsilon_t}{L},\qquad l=1,\dots,L.
\end{equation}
After clipping, client $C_i$ injects independent Laplace noise into each layer's summed gradient and averages by the local sample size:
\begin{equation}
\widetilde{G}_{t,i}^{(l)}
=
\frac{1}{N_i}\Biggl(G_{t,i}^{(l)}+\operatorname{Lap}\!\Biggl(0,\frac{\Delta_{t,i}^{(l)}}{\epsilon_t^{(l)}}\Biggr)\Biggr).
\end{equation}
Conditioned on $\Delta_{t,i}^{(l)}$, this is the standard Laplace mechanism with sensitivity $\Delta_{t,i}^{(l)}$ and privacy parameter $\epsilon_t^{(l)}$, and therefore satisfies $\epsilon_t^{(l)}$-local differential privacy for layer $l$ at round $t$.

The client concatenates all layer-wise noisy updates into a single vector $\widetilde{\mathbf{g}}_t$, encodes it by fixed precision scaling factor $\mathrm{SF}=10^{d_n}$,
\begin{equation}
\widetilde{\mathbf{g}}_{t,\mathrm{encoded}}=\operatorname{round}(\widetilde{\mathbf{g}}_t\cdot\mathrm{SF}),
\end{equation}
and secret-shares the encoded vector among the $m$ intermediate servers. Each intermediate server sums the received shares and sends the partial aggregate to the PS. The PS reconstructs
\begin{equation}
\nabla\theta_{\mathrm{agg}}
=
\Biggl(\sum_{j=1}^{m}\nabla\theta_{\mathrm{agg},j}\Biggr)\bmod p,
\end{equation}
decodes by dividing by $\mathrm{SF}$, and updates the global model by
\begin{equation}
\theta_{t+1}=\theta_t-\eta\cdot\frac{N_i}{N}\cdot\nabla\theta_{\mathrm{agg}}.
\end{equation}
The complete procedure is summarized in Algorithm~\ref{alg:ddp-sa-adaptive}. Relative to static DDP-SA, only the clipping and noise calibration steps change; the encoding and secure aggregation workflow remains unchanged.

\begin{algorithm}[!t]
\caption{DDP-SA-adaptive}
\label{alg:ddp-sa-adaptive}
\textbf{Input:} the set of clients $C = \{C_1, C_2, \ldots, C_n\}$, the number of training rounds $T$, the global model parameters $\theta = \{\theta^{(1)}, \ldots, \theta^{(L)}\}$, the set of intermediate servers $S = \{S_1, S_2, \ldots, S_m\}$, the total privacy budget $\epsilon_{\text{total}}$, the number of samples $N_i$ for client $C_i$, the learning rate $\eta$, the total number of samples $N$ all clients posses, the large prime number $p$, the loss function $\mathcal{L}$, the gradient $\nabla_{\theta_t^{(l)}} \mathcal{L}\left(\theta_t, x_j\right)$ of the loss function $\mathcal{L}$ with respect to the model parameters $\theta_t^{(l)}$ of layer $l$, the fixed precision scaling factor $\text{SF}$, the number of decimal places $d_n$ you want to preserve \newline
\textbf{Output:} the trained global model $\theta_T$
\begin{algorithmic}[1]
\For{each round $t=0,1,\dots,T-1$}
    \State PS broadcasts current model parameters $\theta_t$ to all clients
    \State $\epsilon_t \leftarrow \epsilon_{\text{total}} / T$ \Comment{uniform per-round allocation}
    \For{each client $C_i$ in parallel}
        \For{each layer $l=1,\dots,L$}
            \State $\nabla\theta^{(l)}\leftarrow 0$, $\mathcal{G}_{t,i}^{(l)}\leftarrow\emptyset$
            \For{each sample $x_j\in D_i$}
                \State $\mathbf{g}_t^{(l)}(x_j)\leftarrow\nabla_{\theta_t^{(l)}}\mathcal{L}(\theta_t,x_j)$
                \State $\mathcal{G}_{t,i}^{(l)}\leftarrow\mathcal{G}_{t,i}^{(l)}\cup\{\|\mathbf{g}_t^{(l)}(x_j)\|_1\}$
            \EndFor
            \State $\Delta_{t,i}^{(l)}\leftarrow\operatorname{median}(\mathcal{G}_{t,i}^{(l)})$
            \For{each sample $x_j\in D_i$}
                \State $\overline{\mathbf{g}}_t^{(l)}(x_j)\leftarrow\mathbf{g}_t^{(l)}(x_j)$
                \Statex \hspace{\algorithmicindent}$/\max\left(1,\frac{\left\|\mathbf{g}_t^{(l)}\left(x_j\right)\right\|_1}{\Delta_{t,i}^{(l)}}\right)$
                \State $\nabla\theta^{(l)}\leftarrow\nabla\theta^{(l)}+\overline{\mathbf{g}}_t^{(l)}(x_j)$
            \EndFor
            \State $\epsilon_t^{(l)} \leftarrow \epsilon_t / L$ \Comment{uniform per-layer allocation}
            \State $\tilde{\mathbf{g}}_t^{(l)} \leftarrow \frac{1}{N_i}\left(\nabla \theta^{(l)} + \text{Lap}\left(0, \frac{\Delta_{t,i}^{(l)}}{\epsilon_t^{(l)}}\right)\right)$
        \EndFor
        \State Concatenate all $\tilde{\mathbf{g}}_t^{(l)}$ into a single vector $\tilde{\mathbf{g}}_t$
        \State $\mathrm{SF}\leftarrow 10^{d_n}$
        \State $\tilde{\mathbf{g}}_{t,\mathrm{encoded}}\leftarrow\operatorname{round}(\tilde{\mathbf{g}}_t\cdot\mathrm{SF})$
        \State $\mathrm{shares}\leftarrow C_i.\mathrm{secret\_share}(\tilde{\mathbf{g}}_{t,\mathrm{encoded}},S)$
        \For{each share $\mathrm{shares}[j]$}
            \State send $\mathrm{shares}[j]$ to $S_j$
        \EndFor
    \EndFor
    \For{each server $S_j$ in parallel}
        \State $\nabla\theta_{\mathrm{agg},j}\leftarrow$sum of shares received at $S_j$
        \State send $\nabla\theta_{\mathrm{agg},j}$ to the PS
    \EndFor
    \State $\nabla\theta_{\mathrm{agg}}\leftarrow\bigl(\sum_{j=1}^{m}\nabla\theta_{\mathrm{agg},j}\bigr)\bmod p$
    \State $\nabla \theta_{\text{agg}} \leftarrow \frac{\nabla \theta_{\text{agg}}}{\text{SF}}$
    \State $\theta _ { t + 1 } \leftarrow \theta _ { t } - \eta \cdot \frac { N _ { i } } { N } \cdot \nabla \theta _ { \text {agg} }$
\EndFor
\State \Return $\theta_T$
\end{algorithmic}
\end{algorithm}

\subsection{Privacy Analysis}
Across layers within one round, basic composition yields at most $\epsilon_t$-DP. Over $T$ rounds, the overall local mechanism inherits the same multi-round privacy bound as static DDP-SA under the chosen composition theorem~\cite{dwork2014algorithmic,wei2026ddpsa}. Secure aggregation is deterministic post-processing of already privatized updates and therefore introduces no additional privacy loss, while still preventing the PS from observing any individual client update.

\section{Performance Evaluation}
This section evaluates DDP-SA-adaptive against static DDP-SA and a No-Private baseline in terms of efficiency, accuracy, privacy, convergence, clipping norms, and noise magnitude.

\subsection{Experimental Setup}
Experiments were implemented in Python using PyTorch~1.4.0 and PySyft~0.2.9 on GitHub Codespaces (16 cores, 64GB RAM, and 128GB storage). We generated a $10000 \times 2$ array whose
samples were drawn from a uniform distribution and constructed the labels as $y=x_1+x_2+1$, thereby yielding a federated linear regression task. The data were split into training, validation, and test sets in a ratio of $60\%$, $20\%$, and $20\%$, respectively, and the training set was partitioned evenly across clients under an IID setting. The model was a two-layer neural network with two input neurons and one output neuron. The per-round, per-layer privacy budget was set to $\epsilon_t^{(l)}=0.1$, and the number of decimal places was set to $d_n=10$. No-Private used SGD with a learning rate of $0.1$, whereas DDP-SA and DDP-SA-adaptive used Adam with a learning rate of $0.001$. Static DDP-SA employed a fixed clipping threshold, whereas DDP-SA-adaptive used the round-wise, layer-wise median-based clipping method described in Algorithm~\ref{alg:ddp-sa-adaptive}. Due to the randomness of DP noise, the reported results were averaged over multiple runs. We compared DDP-SA-adaptive with static DDP-SA under the same secure aggregation architecture.

\subsection{Efficiency Analysis}
We measure communication cost by the number of communication rounds until convergence and by the number of parameters uploaded per client per round, and measure computation cost by total training time and average per-round time. Table~\ref{tab:efficiency} summarizes the results. No-Private converges in 2082 rounds, while DDP-SA and DDP-SA-adaptive require 2436 and 2270 rounds, respectively. Thus, DDP-SA-adaptive reduces the round count of static DDP-SA by $6.81\%$. Both DDP-SA and DDP-SA-adaptive upload 9 parameters per client when $m=3$, compared with 3 for No-Private, because each encoded coordinate is split into three secret shares. In computation, No-Private, DDP-SA, and DDP-SA-adaptive spend 112 minutes, 203 minutes, and 164 minutes in total, and 6.46 seconds, 10.00 seconds, and 8.67 seconds per round on average. Relative to static DDP-SA, the adaptive mechanism therefore reduces total time by $19.21\%$ and average per-round time by $13.33\%$. Hence, the DDP-SA-adaptive mechanism significantly reduces both communication and computational costs compared to the
original DDP-SA mechanism, leading to improved training efficiency.

\begin{table}[!t]
\setlength{\tabcolsep}{4pt}
\renewcommand{\arraystretch}{1.3}
\caption{Efficiency Comparison under Different Defensive Mechanisms}
\label{tab:efficiency}
\centering
\begin{tabular}{|c|c|c|c|}
\hline
\textbf{Metric} & \textbf{No-Private} & \textbf{DDP-SA} & \textbf{DDP-SA-adaptive} \\
\hline
Communication Rounds & 2082 & 2436 & 2270 \\
Parameters Uploaded  & 3    & 9    & 9    \\
Total Time (min)     & 112  & 203  & 164  \\
Avg Time (s)         & 6.46 & 10.00 & 8.67 \\
\hline
\end{tabular}
\end{table}

\subsection{Accuracy Analysis}
Table~\ref{tab:accuracy} report test loss and test $\mathrm{R}^{2}$. No-Private attains near-zero test loss ($10^{-12}$) and $\mathrm{R}^{2}=0.9999$. Static DDP-SA yields test loss $0.0055$ and $\mathrm{R}^{2}=0.9666$, whereas DDP-SA-adaptive reduces the test loss to $6.9354\times 10^{-5}$ ($98.74\%$ lower) and increases $\mathrm{R}^{2}$ to $0.9996$ ($3.41\%$ higher), approaching the Non-Private baseline. Hence, the DDP-SA-adaptive mechanism significantly reduces test loss and increases test $\mathrm{R}^{2}$ compared to the original DDP-SA mechanism, leading to improved model accuracy.

\begin{table}[!t]
\renewcommand{\arraystretch}{1.3}
\caption{Accuracy Comparison under Different Defensive Mechanisms}
\label{tab:accuracy}
\centering
\begin{tabular}{|c|c|c|c|}
\hline
\textbf{Metric} & \textbf{No-Private} & \textbf{DDP-SA} & \textbf{DDP-SA-adaptive} \\
\hline
Test Loss & $10^{-12}$ & 0.0055 & $6.9354\times 10^{-5}$ \\
Test $\mathrm{R}^{2}$ & 0.9999 & 0.9666 & 0.9996 \\
\hline
\end{tabular}
\end{table}

\subsection{Analysis of Privacy Protection Strength}
\begin{figure}[!t]
\centering
\includegraphics[width=3.2in]{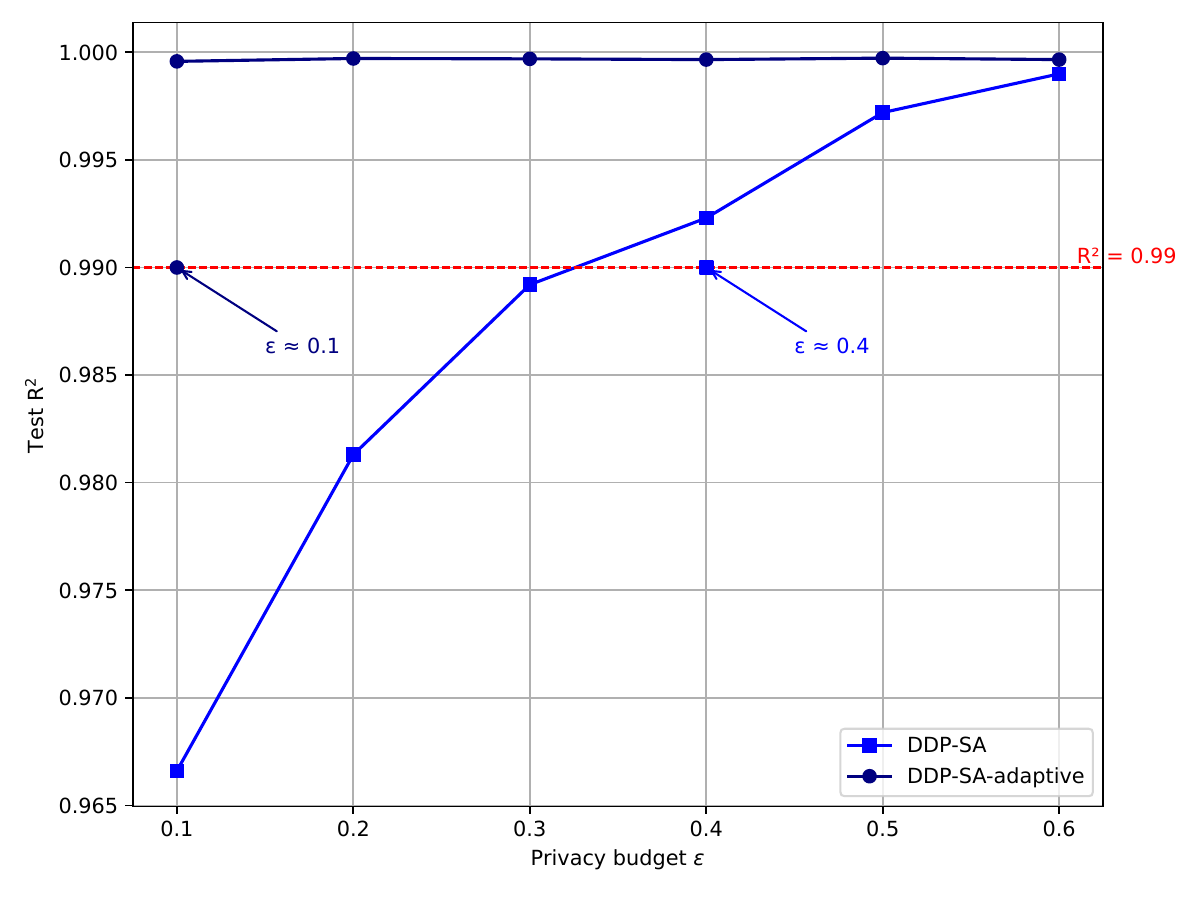}
\caption{Test $\mathrm{R}^{2}$ versus privacy budget $\epsilon$. The red dashed line marks $\mathrm{R}^{2}=0.99$.}
\label{fig:r2-eps}
\end{figure}

Fig.~\ref{fig:r2-eps} plots test $\mathrm{R}^{2}$ against the privacy budget $\epsilon\in[0.1,0.6]$. Smaller $\epsilon$ corresponds to stronger privacy protection. DDP-SA-adaptive remains close to $\mathrm{R}^{2}\approx 1$ across the entire range, whereas static DDP-SA degrades markedly under tight budgets. To reach the target $\mathrm{R}^{2}=0.99$, static DDP-SA requires $\epsilon\approx 0.4$, while DDP-SA-adaptive already meets the target at $\epsilon\approx 0.1$. Hence, for the same accuracy goal, the adaptive mechanism operates under a substantially smaller privacy budget, thus providing stronger privacy protection and achieving stronger privacy guarantees.

\begin{figure}[t!]
\centering
\includegraphics[width=3.2in]{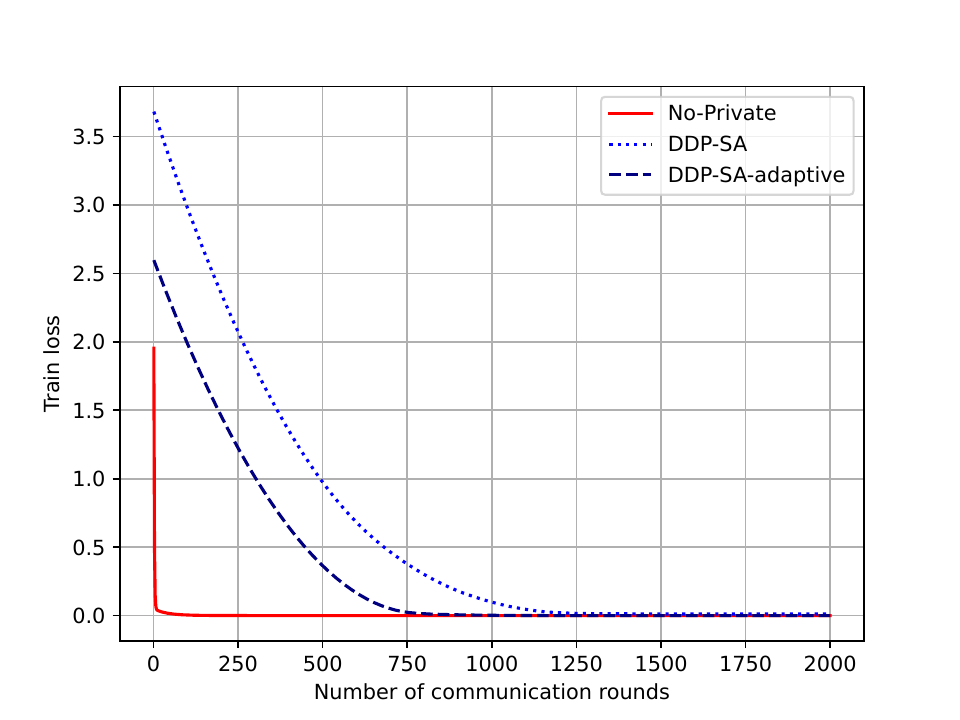}
\caption{Train loss under different number of communication rounds.}
\label{fig:train_loss-rounds}
\end{figure}

\begin{figure}[t!]
\centering
\includegraphics[width=3.2in]{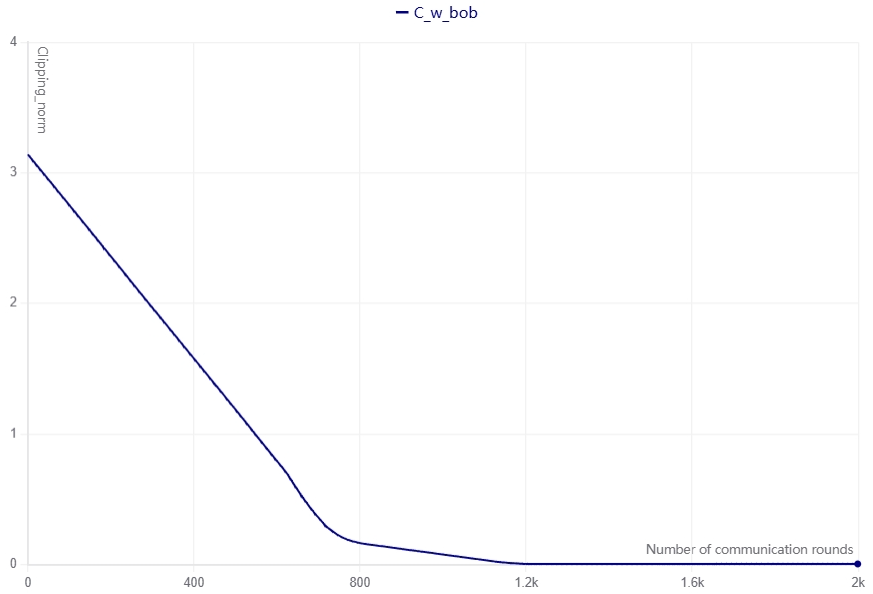}
\caption{Clipping norm of the gradients of the Bob's weights vs number of communication rounds.}
\label{fig:clipping_norm-rounds}
\end{figure}

\begin{figure}[t!]
\centering
\includegraphics[width=3.2in]{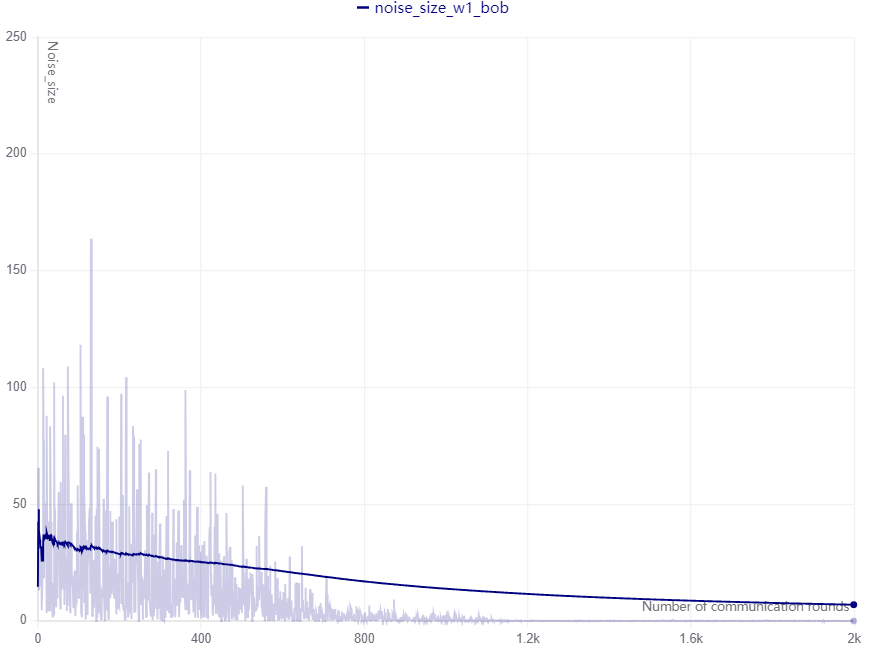}
\caption{Noise magnitude added to the gradient of the weight $w_1$ of Bob versus communication rounds. Shaded curves show raw noise; navy curves show EMA-smoothed trends.}
\label{fig:noise-w1-bob}
\end{figure}

\begin{figure}[t!]
\centering
\includegraphics[width=3.2in]{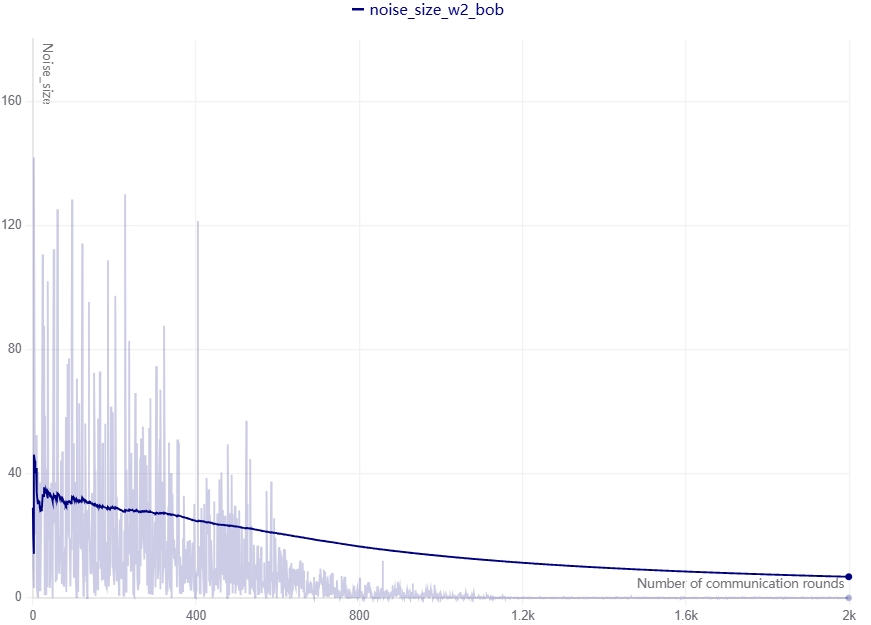}
\caption{Noise magnitude added to the gradient of the weight $w_2$ of Bob versus communication rounds. Shaded curves show raw noise; navy curves show EMA-smoothed trends.}
\label{fig:noise-w2-bob}
\end{figure}

\subsection{Analysis of Convergence, Clipping Norm, and Noise Magnitude}
We further examine why adaptation improves efficiency and accuracy. Practical rounds to convergence are 1041 for No-Private, 1218 for DDP-SA, and 1135 for DDP-SA-adaptive, and the adaptive curve also attains a lower train loss at convergence, as shown in Fig.~\ref{fig:train_loss-rounds}. Fig.~\ref{fig:clipping_norm-rounds} shows that the median-based clipping norm of client Bob decreases rapidly over rounds as gradient magnitudes contract. Because the Laplace scale is proportional to $\Delta_{t,i}^{(l)}/\epsilon_t^{(l)}$, the injected noise follows the same trend: raw noise samples fluctuate locally, but the EMA-smoothed trajectories (EMA $=0.99$) decline steadily, as illustrated in Fig.~\ref{fig:noise-w1-bob} and Fig.~\ref{fig:noise-w2-bob}. Consequently, late stage updates are less distorted, which accelerates convergence and improves final accuracy without changing the secure aggregation workflow. The reason for the above conclusion is that DDP-SA-adaptive
adopts an adaptive gradient clipping and noise injection mechanism, which uses the median of the norms of the unclipped samples’ gradients in the current training round as the clipping norm.
Over the course of training, the train loss decreases gradually, and the predicted label becomes closer to the true label, leading to a decrease in the norm of each sample’s gradient. As a result, the clipping norm decreases gradually, which causes the required noise level to decrease as well. Therefore, DDP-SA-adaptive requires fewer communication rounds to achieve convergence and achieves higher model accuracy than DDP-SA.

\section{Conclusion}
In this paper, we presented DDP-SA-adaptive, an adaptive gradient clipping and noise injection mechanism for differentially private federated learning with secure aggregation. Building on the DDP-SA pipeline, each client replaces a static clipping threshold with a round-wise, layer-wise median of per-sample gradient norms and calibrates Laplace noise accordingly, while preserving fixed precision encoding and additive secret sharing for secure aggregation.

Extensive experiments on a federated regression task show that compared to the original DDP-SA, the proposed DDP-SA-adaptive significantly improves training efficiency and model accuracy while preserving stronger privacy guarantees. Analyses of clipping norms and noise magnitudes further confirm that both quantities decrease as training progresses, which explains the faster convergence and higher final accuracy. Extending the study to deeper architectures and non-IID FL benchmarks is an important direction for future work.

\bibliographystyle{IEEEtran}
\bibliography{bibliography}

\end{document}